\newcommand{\PT}{{\rm P}_{\!\!\scriptscriptstyle\rm T}}
\newcommand{\ET}{{\rm E}_{\scriptscriptstyle\rm T}}
\newcommand{\MET}{\mbox{$\raisebox{.3ex}{$\not$}\ET$}}
\newcommand{\ppbar}{p\bar{p}}
\newcommand{\ttbar}{t\bar{t}}
\newcommand{\bbbar}{b\bar{b}}
\newcommand{\ccbar}{c\bar{c}}
\newcommand{\ipb}{ {\rm pb}^{-1} }
\def \gev {{\rm GeV/c^2}}
\def\mp{${\pm}$}
\documentstyle[aps,epsf]{revtex}

\begin{document}
\draft
\preprint{Fermilab-Pub 97/286-E}
\title{Measurement of the $\ttbar$ Production Cross Section in $\ppbar$
Collisions at $\sqrt{s}=1.8$~TeV}
\author{\font\eightit=cmti8
\def\r#1{\ignorespaces $^{#1}$}
\hfilneg
\begin{sloppypar}
\noindent
F.~Abe,\r {17} H.~Akimoto,\r {39}
A.~Akopian,\r {31} M.~G.~Albrow,\r 7 A.~Amadon,\r 5 S.~R.~Amendolia,\r {27} 
D.~Amidei,\r {20} J.~Antos,\r {33} S.~Aota,\r {37}
T.~Arisawa,\r {39}
% ...
\end{sloppypar}
\vskip .026in
\begin{center}
(CDF Collaboration)
\end{center}

\vskip .026in
\begin{center}
\r 1  {\eightit Argonne National Laboratory, Argonne, Illinois 60439} \\
\r 2  {\eightit Istituto Nazionale di Fisica Nucleare, University of Bologna,
I-40127 Bologna, Italy} \\
\r 3  {\eightit Brandeis University, Waltham, Massachusetts 02254} \\
\r 4  {\eightit University of California at Los Angeles, Los 
Angeles, California  90024} \\  
\r 5  {\eightit University of Chicago, Chicago, Illinois 60637} \\
\r 6  {\eightit Duke University, Durham, North Carolina  27708} \\
\r 7  {\eightit Fermi National Accelerator Laboratory, Batavia, Illinois 
60510} \\
\r 8  {\eightit University of Florida, Gainesville, FL  32611} \\
\r 9  {\eightit Laboratori Nazionali di Frascati, Istituto Nazionale di Fisica
               Nucleare, I-00044 Frascati, Italy} \\
\r {10} {\eightit University of Geneva, CH-1211 Geneva 4, Switzerland} \\
\r {11} {\eightit Harvard University, Cambridge, Massachusetts 02138} \\
\r {12} {\eightit Hiroshima University, Higashi-Hiroshima 724, Japan} \\
\r {13} {\eightit University of Illinois, Urbana, Illinois 61801} \\
\r {14} {\eightit Institute of Particle Physics, McGill University, Montreal 
H3A 2T8, and University of Toronto,\\ Toronto M5S 1A7, Canada} \\
\r {15} {\eightit The Johns Hopkins University, Baltimore, Maryland 21218} \\
\r {16} {\eightit Institut f\"{u}r Experimentelle Kernphysik, 
Universit\"{a}t Karlsruhe, 76128 Karlsruhe, Germany} \\
\r {17} {\eightit National Laboratory for High Energy Physics (KEK), Tsukuba, 
Ibaraki 305, Japan} \\
\r {18} {\eightit Ernest Orlando Lawrence Berkeley National Laboratory, 
Berkeley, California 94720} \\
\r {19} {\eightit Massachusetts Institute of Technology, Cambridge,
Massachusetts  02139} \\   
\r {20} {\eightit University of Michigan, Ann Arbor, Michigan 48109} \\
\r {21} {\eightit Michigan State University, East Lansing, Michigan  48824} \\
\r {22} {\eightit University of New Mexico, Albuquerque, New Mexico 87131} \\
\r {23} {\eightit The Ohio State University, Columbus, OH 43210} \\
\r {24} {\eightit Osaka City University, Osaka 588, Japan} \\
\r {25} {\eightit Universita di Padova, Istituto Nazionale di Fisica 
          Nucleare, Sezione di Padova, I-36132 Padova, Italy} \\
\r {26} {\eightit University of Pennsylvania, Philadelphia, 
        Pennsylvania 19104} \\   
\r {27} {\eightit Istituto Nazionale di Fisica Nucleare, University and Scuola
               Normale Superiore of Pisa, I-56100 Pisa, Italy} \\
\r {28} {\eightit University of Pittsburgh, Pittsburgh, Pennsylvania 15260} \\
\r {29} {\eightit Purdue University, West Lafayette, Indiana 47907} \\
\r {30} {\eightit University of Rochester, Rochester, New York 14627} \\
\r {31} {\eightit Rockefeller University, New York, New York 10021} \\
\r {32} {\eightit Rutgers University, Piscataway, New Jersey 08855} \\
\r {33} {\eightit Academia Sinica, Taipei, Taiwan 11530, Republic of China} \\
\r {34} {\eightit Texas A\&M University, College Station, Texas 77843} \\
\r {35} {\eightit Texas Tech University, Lubbock, Texas 79409} \\
\r {36} {\eightit Istituto Nazionale di Fisica Nucleare, University of Trieste/
Udine, Italy} \\
\r {37} {\eightit University of Tsukuba, Tsukuba, Ibaraki 315, Japan} \\
\r {38} {\eightit Tufts University, Medford, Massachusetts 02155} \\
\r {39} {\eightit Waseda University, Tokyo 169, Japan} \\
\r {40} {\eightit University of Wisconsin, Madison, Wisconsin 53706} \\
\r {41} {\eightit Yale University, New Haven, Connecticut 06520} \\
\end{center}

}
\date{\today}
\maketitle

\begin{abstract}
We present a measurement of the $\ttbar$ production cross section in
$\ppbar$ collisions at $\sqrt{s}=1.8$~TeV using an integrated luminosity
of 109 $\ipb$ collected with the Collider Detector at Fermilab.
The measurement uses $\ttbar$ decays into 
final states which contain one or two high transverse momentum leptons and 
multiple jets, and final states which contain only jets.
Using acceptances appropriate for
 a top quark mass of 175$~\gev$, we find $\sigma_{\ttbar}=7.6^{+1.8}_{-1.5}$ pb.
\end{abstract}
The measurement of the $\ppbar\rightarrow\ttbar~X$ production cross section
presents a test of both the production and decay mechanisms of the standard
model.
Recent calculations~\cite{mlm_xsec} based on Quantum Chromodynamics (QCD)
have led to predictions for
the cross section with a theoretical uncertainty of less than 15\%.  
A measurement that is significantly different from the predicted value can
signal either non-standard model production, for instance the decay of a heavy
resonance into $\ttbar$ pairs, or a non-standard model decay mechanism such
as the decay into supersymmetric particles~\cite{Lane}.  In the latter case
it is of particular interest to measure the cross section into different
final states, because an unexpected decay mode of the top quark will modify
the expected branching fractions.
The $\ttbar$ production cross section has been
measured before by both the Collider Detector at Fermilab (CDF) and D0
collaborations~\cite{evidence_prd,obs_prl,D0_prl}. 

The standard model predicts that the top quark will decay nearly 100\% of
the time to $Wb$.   The $W$ boson can then 
decay to either a pair of quarks, or
a lepton neutrino ($\ell\nu$) pair. 
We categorize the decays of $\ttbar$ pairs by the decays of
the two $W$ bosons as either lepton+jets, dilepton, or all-hadronic.
The dilepton and all-hadronic analyses are described  
elsewhere~\cite{dil_prl,had_prl}.
We now have nearly twice as much data as reported in~\cite{obs_prl}.
With 
improved measurements of acceptances and backgrounds, and by combining all the
decay modes, we measure the cross section with better than twice the
precision of our previous measurement.

The data presented here represent the entire data set accumulated between
1992 and 1995 with the CDF detector, and corresponds to
an integrated luminosity of 109$\pm7~\ipb$
 (19~$\ipb$  from the 1992-93 run and 90~$\ipb$ from the 
1994-95 run)~\cite{wz_xsec}.  

The CDF detector consists of a magnetic spectrometer surrounded by calorimeters
and muon chambers.  A four-layer silicon vertex detector (SVX),
located immediately outside the beam pipe, provides precise track reconstruction
in the plane transverse to the beam and is used to identify secondary vertices
from $b$ and $c$ quark decays.
A detailed description of the detector can be found
elsewhere~\cite{evidence_prd,NIM}.

The electron, muon, and multi-jet events used in this analysis were selected by
a three-level trigger.  Lepton samples were acquired with inclusive electron and
muon triggers requiring $\PT$(lepton)$ > 18$ GeV/c. 
A missing transverse energy~\cite{evidence_prd}, $\MET$,
trigger was also used in
order to recover events lost due to small inefficiencies in the inclusive 
lepton triggers.  

Decays of $\ttbar$ pairs into lepton+jets are characterized by a single 
high-$\PT$ lepton, missing transverse energy from the $W\rightarrow\ell\nu$
decay, plus four jets, two from the hadronically decaying $W$ boson and two
from the $b$ quarks from the top decays.
Jets are defined using a cone algorithm with $\Delta R=
\sqrt{\Delta\phi^2 +\Delta\eta^2}$=0.4, where $\eta$ is the pseudo-rapidity. 
Jets are counted in this analysis if $\mid\eta\mid <2.0$.
The number of observed jets may decrease due to detector effects or jet
overlap, or increase as a result of multiple interactions or the presence
of gluon radiation.
In the lepton+jets channel, events with three or more jets with measured
$\ET >$15 GeV define the $\ttbar$ signal region.

The data sample for the lepton+jets analysis is a subset of a sample of
high-$\PT$ inclusive lepton events that contain either an isolated electron with
$\ET >$ 20 GeV or an isolated muon with $\PT >$ 20 GeV/c  in the central region
($\mid\eta\mid <$1.0).  Events that contain a second same
flavor lepton of opposite charge are removed as $Z$ boson candidates 
if the reconstructed $ee$ or $\mu\mu$ invariant mass is between 75 and 105
GeV/c$^2$.   If a candidate high-$\ET$ photon~\cite{pho_prl} is present, the
three-body mass is used to remove radiative $Z$ candidates. 
An inclusive $W$
boson sample is selected from the inclusive lepton sample by requiring $\MET
~>$ 20 GeV and that the lepton be isolated from any jet activity. 
For the latter we define isolation, $I$,
as the transverse energy in a cone of $\Delta R=$ 0.4 centered on the lepton,
but excluding the lepton energy, divided by the $\ET~(\PT)$ of the electron
(muon), and require $I~<$ 0.1 .
 Furthermore, 
the event must not be accepted
as a  dilepton candidate~\cite{dil_prl}.  

In order to separate $\ttbar$ events in the lepton+jets channel from the large
$W$+jets background, we require that one of the jets be identified as 
a $b$ jet candidate.
Identification of $b$ jets is done either by reconstructing secondary
vertices from $b$-quark decay using the SVX (SVX tagging), or by identifying
an additional lepton from a semileptonic $b$ decay 
(SLT tagging).  The SVX and SLT tagging algorithms are described in
Ref.~\cite{obs_prl}.

The efficiency for tagging a $b$ quark from a top decay is determined from 
$\ttbar$ Monte Carlo data together with a detailed simulation of the detector,
which includes the effects of local track density on the track finding
efficiency. The systematic uncertainty due to the tracking efficiency modeling
is determined by comparing data and Monte Carlo tracking efficiencies and
multiplicity distributions as a function of jet $\ET$ in inclusive electron and
muon samples, which are enriched in $b$ decays.  The efficiency for tagging at
least one $b$ quark in a $\ttbar$ event with $\geq$ 3 jets is found to be
(39$\pm$3)\%.  Of this 39\%, a factor of 67\% comes from the fiducial acceptance
of the SVX.

The SLT algorithm identifies both muons and electrons with $\PT >2.0$ GeV/c to
$\mid\eta\mid=1.0$.  The efficiency of this algorithm, as a function of lepton
$\PT$, is measured with photon conversion and $J/\psi \rightarrow \mu\mu$ data,
and applied to Monte Carlo $\ttbar$ events.  The probability of finding an
additional $e$ or $\mu$ from a $b$ quark decay in a $\ttbar$ event with $\geq$ 3
jets is  (18 $\pm$ 2)\%.  

In the $\ttbar$ signal region of $W+\geq 3$ jets, there are 
34 SVX-tagged events containing a total of 42 SVX tags, and 40 SLT-tagged events
containing a total of 44 SLT tags.
Of these, 11 events are tagged by both
the SVX and SLT algorithms. 

The acceptance for identifying $\ttbar$ events in the lepton+jets mode
is calculated
from a combination of data and {\sc pythia}~\cite{pythia} and 
{\sc herwig}~\cite{herwig} $\ttbar$ Monte Carlo samples.  
We use a top mass of 175 $\gev$~\cite{Tipton} when
evaluating the acceptance.  
The total acceptance, including the branching fraction,
is calculated as the product of the kinematic (including lepton identification)
and geometric acceptances, the
trigger efficiency and the tagging efficiency.  We measure
these efficiencies as described in Ref.~\cite{evidence_prd}, and average over
the two running periods.
For the lepton+jets analysis, the product of the geometric and kinematic
acceptance is (10.4$\pm$1.0)\%, and the trigger efficiency is
(90$\pm$7)\%.  These factors are common between the
SVX and SLT analyses. Combining with the respective tagging efficiencies
gives a total SVX acceptance of (3.7$\pm$0.5)\% and a
total SLT acceptance of (1.7$\pm$0.3)\%.

The systematic uncertainties on the geometric and kinematic acceptances come
from the following:  
the jet energy scale ($\pm$5\%), modeling of initial state gluon radiation 
($\pm$2\%), final state gluon radiation ($\pm$5\%), Monte Carlo dependence and
modeling ($\pm$5\%),
detector resolution effects ($\pm$2\%) and
instantaneous luminosity dependence ($\pm$1\%).  
The uncertainties on the
tagging and trigger efficiencies are dominated by the level of agreement 
between data and the Monte Carlo predictions.  


The most important source of background in the SVX-tagged lepton+jets
channel is
inclusive $W$ production in association with jets containing $b$ or $c$ quarks,
eg. $p\bar{p} \rightarrow Wg$ ($g \rightarrow b\bar{b}$).
In addition, there are contributions to the background from mistags
(i.e. tags in jets which contain no true displaced vertices),
and  small contributions
from the following processes: non-$W$(e.g. direct $b\bar{b}$ production), 
single top production, $WW$, $WZ$, and Drell-Yan.

To calculate the background from $W$+heavy flavor events, we use the {\sc
herwig} and {\sc vecbos}\cite{vecbos} Monte Carlo programs to predict, as a
function of jet multiplicity, the fraction of $W+$jet events which are
$W\bbbar$, $W\ccbar$ and $Wc$.  These fractions, and a tagging  efficiency for
each type of event, are applied to the number of $W+$jet events seen in the data
to give an expected background from these sources for each jet multiplicity.
The details of this method can be found in Ref.~\cite{evidence_prd}.

The background from events in the sample that do not contain real $W$
bosons (non-$W$) is calculated from the data by measuring the number of
tags as a function of lepton isolation, $I$, and $\MET$.  The tagging
rate in the low $\MET$, high-$I$ region, where there are essentially no real
$W$ events, is used to predict the contamination in the $W$ signal region of
high $\MET$, low $I$.

To calculate the background from mistags ~\cite{evidence_prd}, 
we assume that the distribution of
reconstructed transverse decay length, L$_{xy}$, from this source is symmetric
about zero.  Secondary vertices with negative L$_{xy}$ (i.e. those
which reconstruct to the opposite side of the primary from the jet direction)
come primarily from reconstruction errors in light quark jets. 
We parametrize the negative L$_{xy}$ distribution measured in
generic jet data as a function of jet $\ET$, $\eta$, and the number of SVX
tracks in the jet.  This parametrization is applied to the $W$+jets data to
predict the number of mistags observed. 

The single top background is determined by measuring the acceptance for
$W$* and $W$-gluon production using the {\sc pythia} and {\sc herwig}
 Monte Carlo programs,
and using the latest theoretical cross sections~\cite{scottim}.
The remaining, relatively small, backgrounds ($WW$, $WZ$, $Z\rightarrow
\tau\tau$) are derived from Monte Carlo predictions.  

The individual
components of the background and their totals
are shown in Table~\ref{wjets}.  
In addition to the signal region of 3 or more jets, we show the
predicted number of tags in events with 1 and 2 jets as a check
of our calculation.
An iterative correction, to account for the $\ttbar$ content of the $W$+jet
events, is applied to those backgrounds that are calculated as a fraction of
the observed number of these events~\cite{evidence_prd}.
The corrected background in the signal region is 9.2$\pm$1.5 tagged events.
We observe 34 tagged events, resulting in a cross section of 
6.2$^{+2.1}_{-1.7}$~pb.

The background to SLT-tagged events is dominated by $W$ events with hadrons
misidentified as leptons (including decays in flight), with electrons from
unidentified photon conversions, or with real heavy flavor jets ($W\bbbar$,
$W\ccbar$).  These backgrounds are calculated by measuring the fraction of tags
per track in a generic jet sample as a function of the track $\PT$.  These
fractions are applied to tracks in the $W$+jet events to estimate the
background from the above sources.  Smaller backgrounds are, in order of
importance in the signal region, $WW$ and $WZ$, non-$W$, $Z\rightarrow\tau\tau$,
single top, $W$c, and Drell-Yan production. The results of the background
calculation and the number of tags observed in the data are shown in
Table~\ref{wjets}.  
In the signal region of $W + \geq$3 jets, the
background prediction is 22.6 $\pm$ 2.8 tagged events. We observe 40 SLT tagged
events, resulting in a cross section of 9.2 $^{+4.3}_{-3.6}$~pb.


\begin{table}
\begin{center}
\begin{tabular}{|l|c|c|c|c|} \hline      
		  & $W$+1 Jet & $W$+2 Jets & $W$+3 Jets  & $W+\geq$4 Jets \\
\hline
Events before tagging    & 10 716    & 1663      & 254         & 70  	 	 
\\ 
\hline
SVX tagged events & 70        & 45         & 18          & 16		  \\
\hline
$Wb\bar{b} + Wc\bar{c} $
	         &19.3 \mp 6.7 & 9.7 \mp 2.4 & 2.3 \mp 0.6 & 0.85 \mp 0.24 \\ 
Non-$W$	         &7.7 \mp 3.0 & 4.0 \mp 1.5  & 1.4 \mp 0.5  & 0.77 \mp 0.33 \\ 
Mistags	         &20.9 \mp 6.3 & 7.2 \mp 2.1 & 1.7 \mp 0.5  & 0.63 \mp 0.22 \\ 
Single top       &1.3 \mp 0.4 & 2.8 \mp 0.7  & 1.0 \mp 0.4 & 0.29 \mp 0.14 \\
Other (incl. $Wc$)&21.5 \mp 5.2 & 7.4 \mp 1.5 & 1.3 \mp 0.2 & 0.39 \mp 0.08 \\
\hline
Uncorrected bkgnd total 
		& 71 \mp 11 & 31 \mp  4   & 7.7 \mp 1.1  & 2.9 \mp 0.5\\ 
Corrected bkgnd total 
		& 71 \mp 11 & 31 \mp  4   & 7.2  \mp 1.1  & 2.0 \mp 0.4\\ 
\hline
SVX tagged $\ttbar$ expected 
		  & 1.0$\pm$ 0.3 & 6.9$\pm$2.1 & 13.3$\pm$3.6 & 17.7$\pm$4.7 \\
\hline\hline
SLT tagged events  
	  & 241           & 78           & 25           &  15 \\
\hline
Mistags+$W\bbbar$+$W\ccbar$	  & 235   \mp 21  & 66.6 \mp 6.1 
				  & 15.1 \mp 1.4 & 6.8  \mp 0.7 \\
Single top& 0.9   \mp 0.3 & 1.5  \mp 0.5 & 0.6  \mp 0.3 & 0.2  \mp 0.1  \\
Other & 33.1 \mp 10.6 & 9.6 \mp 3.0 & 1.2 \mp 1.4 & 0.6 \mp 0.6 \\
\hline
Uncorrected bkgnd total
         & 269$\pm$23     & 77.7$\pm$6.6 & 16.9$\pm$2.0 & 7.6$\pm$0.9\\
Corrected bkgnd total
         & 269$\pm$23     & 77.7$\pm$6.6 & 15.9$\pm$2.0 & 6.7$\pm$0.8\\
 \hline 
SLT tagged $\ttbar$ expected
	& 0.8$\pm$0.2 & 3.8$\pm$1.2 & 6.6$\pm$1.7 & 7.7 $\pm$2.1 \\ \hline
\end{tabular}
\end{center}
\caption{Summary of event yields from the lepton+jets analyses.  The expected
$\ttbar$ contribution is calculated using the measured combined cross section
from this paper.}
\label{wjets}
\end{table}


Our best measurement of the $\ttbar$ cross section comes from combining the
results of the lepton+jets analyses  with the dilepton and
all-hadronic analyses~\cite{dil_prl,had_prl}.  The results of the individual
analyses are summarized in Table~\ref{acc-tab}.  The dilepton analysis finds 9
candidate events, with an expected background of 2.4$\pm$0.5.  The all-hadronic
analysis has two parallel paths, one which requires a single SVX tagged jet plus
kinematic cuts to isolate $\ttbar$, and a second which requires two SVX tagged
jets, but no additional kinematic cuts.  The single tag analysis identifies
187 candidate events with an expected background of 142$\pm$12 events, while the
double tag analysis identifies 157 candidates and predicts 120$\pm18$ background
events.  There are 34 candidate events in common between the two analyses.
The dilepton, lepton+jets, and all-hadronic data samples are
exclusive sets.

We calculate the $\ttbar$ production cross section from the 
combined results of the dilepton and lepton+jets channels
using the same maximum likelihood technique described in~\cite{evidence_prd}.
The all-hadronic result is added by including the multivariate Gaussian term
described in Ref.~\cite{had_prl}.
The likelihood properly accounts for
correlated systematic uncertainties, such as the uncertainty on the integrated
luminosity,
and the uncertainty on the
lepton+jets geometric and kinematic acceptance, which is common to both the
SVX and SLT analyses.

\begin{table}
\begin{center}
\small
\begin{tabular}{|l|c|c|c|c|c|} \hline
 & \multicolumn{2}{c|}{Lepton+Jets} & Dilepton 
& \multicolumn{2}{c|}{All-Hadronic} \\ \hline
Tag & SVX & SLT & not req. & SVX & 2 SVX \\ \hline\hline
$\epsilon_{\rm tag}$ & 0.39$\pm$0.03 & 0.18$\pm$0.02 &$-$ 
& 0.42$\pm$0.04 & 0.11$\pm$0.02 \\
$\epsilon_{geo \cdot kin}$ & 
\multicolumn{2}{c|}{ 0.104$\pm$0.010 } 
& 0.0076$\pm$0.0008 & 0.106$\pm$0.021 & 0.263$\pm$0.045 \\
$\epsilon_{trigger}$ & 
\multicolumn{2}{c|}{$0.90\pm 0.07 $} 
& 0.98$\pm$0.01 & 
\multicolumn{2}{c|}{ 0.998$^{+0.002}_{-0.009}$} \\ \hline
$\epsilon_{\rm total}$ & 0.037$\pm$0.005 & 0.017$\pm$0.003 & 
 0.0074$\pm$0.0008 & 0.044$\pm$0.010 & 0.030$\pm$0.010 \\
\hline
Obs. Events & 34 & 40 & 9 & 187 & 157 \\
Background  & 9.2 \mp 1.5  & 22.6 \mp 2.8 & 2.4 \mp 0.5 & 142 \mp 12 
& 120 \mp 18 \\
\hline
$\sigma_{\ttbar}$ (pb) & 6.2$^{+2.1}_{-1.7}$ & 9.2 $^{+4.3}_{-3.6}$ &
 8.2$^{+4.4}_{-3.4}$ & 9.6$^{+4.4}_{-3.6}$ 
& 11.5$^{+7.7}_{-7.0}$\\
\hline
\end{tabular}
\end{center}
\caption{Summary of acceptance factors and measured cross sections for each
analysis channel.  The acceptances are calculated for a top quark mass of 175
$\gev$.}
\label{acc-tab}
\end{table}
\normalsize

\begin{figure}
\epsfxsize=6.5in
\epsffile[122 180 570 650]{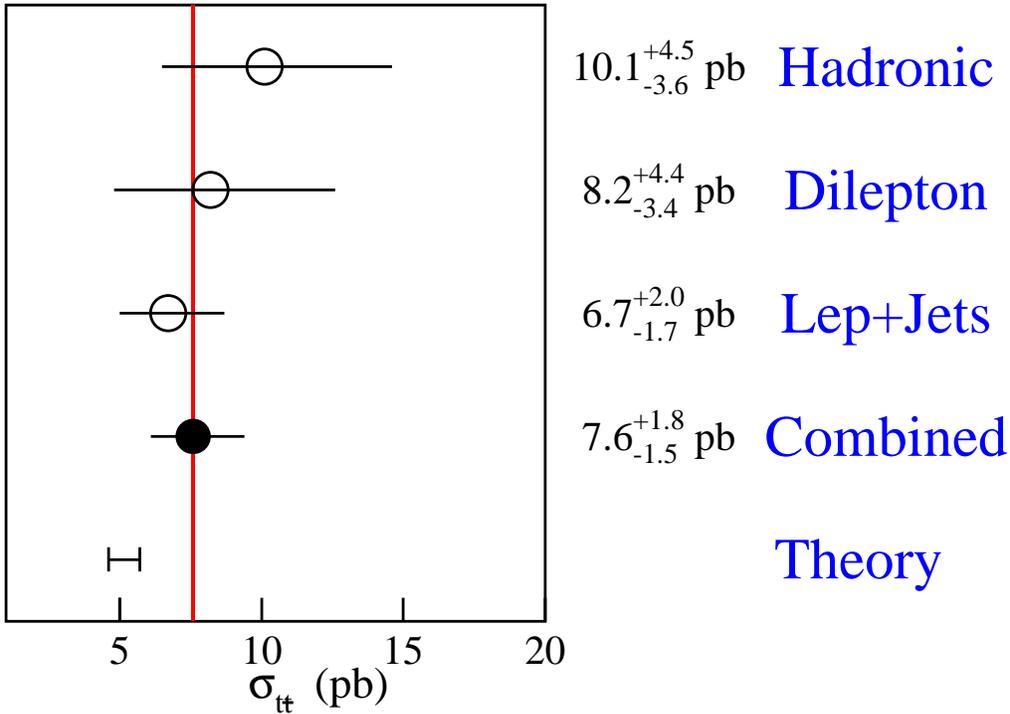}
\caption{Measured $\ttbar$ production cross section at M$_{top}$=175 $\gev$ 
for each of the decay channels and for the combined measurement.  The
lepton+jets cross section is calculated from the SVX and SLT
analyses described in the text.  The
line is drawn through the central value of the combined measurement.
The theory point shows the spread in the central values of the 3 most recent
predictions~\protect\cite{mlm_xsec}.}
\label{xsec_indiv}
\end{figure}

In Figure~\ref{xsec_indiv} we show the results of the cross section calculation
for each $\ttbar$ decay channel, as well as the combined measurement.  
The combined cross section for M$_{top}=175~\gev$ is
7.6$^{+1.8}_{-1.5}$ pb , where the quoted uncertainty includes both statistical
($\pm$1.2 pb) and systematic effects. 
Due to the mass dependence of the acceptances, the calculated cross section
changes by $\pm$10\% for a $\pm$15 GeV/c$^2$ change in top mass from 175
GeV/c$^2$.
Theoretical calculations~\cite{mlm_xsec} range from 4.75 pb to 5.5 pb for
M$_{top}=175~\gev$.  From the ratio of the  measured cross sections in the
dilepton, lepton+jets and all-hadronic channels we can calculate the branching
fraction for a top quark decay to a final state electron or muon. 
Assuming lepton universality and $W$ decay acceptance, the apparent branching
fraction to an electron or muon is 0.188+/-0.048, consistent with the
standard model expectation of $\frac{2}{9}$.  Specifics of possible non-standard
model top decays have not yet been considered.

We thank the Fermilab staff and the technical staffs of the participating
institutions for their vital contributions.  We are grateful for conversations
with T. Stelzer and S. Willenbrock. 
This work is supported by the U.S. Department of Energy
and the National Science Foundation; the Natural Sciences and Engineering
Research Council of Canada; the Istituto Nazionale di Fisica Nucleare of
Italy; the Ministry of Education, Science and Culture of Japan; the National
Science Council of the Republic of China; and the A.P. Sloan Foundation.

%
\end{document}